\documentclass[aps,prl,twocolumn,showpacs]{revtex4}
\begin{document}
\def\bld#1{{\bf #1 }}
 \def\xb{\bar\xi(a,x)}
 \def\gaprox{\mbox{$\,$ 
\raisebox{0.5ex}{$<$}\hspace{-1.7ex}{\raisebox{-0.5ex}{$\sim$ }}$\,$} }
\def\part#1#2{\frac{\partial #1}{\partial #2}}
\def\pder#1#2{{\partial #1/\partial #2}}
\def\rb{\right)}
\def\lb{\left(}
\def\frab#1#2{\left({#1\over#2}\right)}
\def\fra#1#2{{#1\over#2}}
\def\la{\mathrel{\mathchoice {\vcenter{\offinterlineskip\halign{\hfil
$\displaystyle##$\hfil\cr<\cr\sim\cr}}}
{\vcenter{\offinterlineskip\halign{\hfil$\textstyle##$\hfil\cr<\cr\sim\cr}}}
{\vcenter{\offinterlineskip\halign{\hfil$\scriptstyle##$\hfil\cr<\cr\sim\cr}}}
{\vcenter{\offinterlineskip\halign{\hfil$\scriptscriptstyle##$\hfil\cr<\cr\sim\cr}}}}}
\def\ga{\mathrel{\mathchoice {\vcenter{\offinterlineskip\halign{\hfil
$\displaystyle##$\hfil\cr>\cr\sim\cr}}}
{\vcenter{\offinterlineskip\halign{\hfil$\textstyle##$\hfil\cr>\cr\sim\cr}}}
{\vcenter{\offinterlineskip\halign{\hfil$\scriptstyle##$\hfil\cr>\cr\sim\cr}}}
{\vcenter{\offinterlineskip\halign{\hfil$\scriptscriptstyle##$\hfil\cr>\cr\sim\cr}}}}}
\newcommand{\stt}{\small\tt}
\title{Gravitational clustering in Static and Expanding Backgrounds}
\author{T. Padmanabhan}
\email[Email: ]{nabhan@iucaa.ernet.in}
\affiliation{IUCAA, 
Post Bag 4, Ganeshkhind, Pune - 411 007}
\begin{abstract}
A brief summary of several topics in the study of gravitational many body problem is given. The discussion covers both
static backgrounds (applicable to astrophysical systems) as well as  clustering in an expanding background (relevant for cosmology).
\end{abstract}
\maketitle
\section{Introduction }
  
  The statistical mechanics of systems dominated by gravity is of interest both
  from the theoretical and ``practical''  perspectives. Theoretically, this field has close connections
  with areas of condensed matter physics, fluid mechanics, renormalization 
group, etc.  From the practical point of
  view, the ideas find application in different areas of astrophysics and cosmology,
  especially in the study of globular clusters, galaxies and gravitational clustering in the
  expanding universe. [For a review of statistical mechanics of gravitating systems in static background,
  see \cite{tppr}; textbook descriptions are in \cite{textone}; gravitational clustering in cosmology is reviewed  in \cite{tpiran} and
  in the  textbooks \cite{cosmotext}; for a sample of 
   different approaches see \cite{chavanis} and the references cited therein.]
 
\section{Gravitational clustering in static background}
 
To construct the statistical description of a  system of $N$ self gravitating point particles, one should begin
  with the construction of the micro canonical ensemble describing such a system. 
If $g(E)$ is the volume of the constant
  energy surface  $H(p_i, q_i) =E$, then the  entropy  and the temperature of the
  system will be $S(E) = \ln g(E) $ and $T(E)\equiv \beta(E)^{-1} = (\pder{S}{E})^{-1}$.
   (The finiteness of $g$ requires the system
to be confined to a finite volume in space for {\it any} system).

Systems for which a description based on 
  canonical ensemble is possible, the Laplace transform of $g(E)$ with respect to a variable
  $\beta$ will give the partition function $Z(\beta)$. 
  Gravitating systems of interest in astrophysics, however,  cannot be described by a canonical
  ensemble \cite{tppr}, \cite{dlbone}.
  Virial theorem holds for such systems  and we have $(2K+U) =0,$ where 
  $K$ and $U$ are the total kinetic and potential energies of the system.
  This leads to $E=K+U= -K$; since the temperature of the system is proportional
  to the total kinetic energy, the specific heat will be negative: $C_V \equiv (\pder{E}{T})_V 
  \propto (\pder{E}{K}) < 0$.   On the other hand, the specific heat of any system
  described by a canonical ensemble $C_V = \beta^2 \langle(\Delta E)^2\rangle$ will be
  positive definite. Thus, one cannot describe self gravitating systems
  of the kind we are interested in by canonical ensemble.
  
  One can, however,  attempt to find the equilibrium configuration for self gravitating
  systems by maximising the entropy $S(E)$ or the phase volume 
  $g(E)$.
  For a  system of point particles, there is again  no global maximum for entropy \cite{tppr},\cite{textone}. If we move  a small
  number of these particles arbitrarily close to 
  each other, the potential energy of interaction of a pair of these particles, 
  $-Gm_1m_2/r_{12}$, will become arbitrarily high as $r_{12}\to 0$. Transferring some
  of this energy to the rest of the particles, we can increase their kinetic energy without limit.
  This will clearly increase the phase volume occupied by the system (in the momentum space) without bound.
  This argument can be made more formal by dividing the original system into a small,
  compact core, and a large diffuse halo and allowing the core to collapse and transfer the energy to the halo.
  
  The absence of the global maximum for entropy - as argued above - depends on the lack of small scale cutoff. If we assume, instead, that each particle has a  radius $a$, there  will be an upper bound on the 
  amount of energy that can be made available to the rest of the system.
  Further, no real system is completely isolated  and to obtain  a truly isolated system,  we need to confine
  the system  inside a spherical region of radius $R$ with, say, reflecting
  wall. 
  
 The two cut-offs $a$ and $R$ will make the upper bound on the entropy finite, but
 even with  the two cut-offs the formation of a compact
  core and a diffuse halo will still occur, since this is the direction of increasing
  entropy. Particles in the hot diffuse component will permeate the entire spherical
  cavity, bouncing off the walls and having a kinetic energy which is 
  significantly larger than the potential energy. The compact core will exist as 
  a gravitationally bound system with very little kinetic energy. 
  A formal way of understanding this phenomena is based 
  on  the virial theorem \cite{textone}: 
  \begin{equation}
  2T + U = 3PV + \Phi
  \label{modvirial}
  \end{equation}
   for a system with a short distance cut-off
  confined to a sphere of volume $V$, where $P$ is the pressure on the walls and $\Phi$ is the correction to the potential
  energy arising from the short distance cut-off. This equation can be satisfied
  in essentially three different ways. If $T$ and $U$ are significantly higher than
  $3PV$ and $\Phi$, then we have $2T + U \approx 0$ which describes a self gravitating 
  systems in standard virial equilibrium, but not in the state of maximum entropy.
  If $T \gg U$ and $3PV\gg \Phi$, one can have $2T \approx 3PV$ which
  describes an ideal gas with no potential energy confined to a container of
  volume $V$; this will describe the hot diffuse component at late times.
  If  $T \ll U$ and $3PV\ll \Phi$, then one can have $U\approx \Phi$, describing
  the compact potential energy dominated core at late times.
    Such an asymptotic state with two distinct phases  is quite different from what would have
  been expected for systems with only short range interaction. If the gravitating system is put in a heat bath
and the temperature is varied, a sudden phase transition occurs at a critical temperature, leading to the formation
of the two phases  \cite{aaron}, \cite{tppr}.

There are, however, configurations which are {\it local} {\it extrema} of entropy, 
  which are not global maxima. Intuitively, one would have expected
  the distribution of matter in such configuration  to be described by a Boltzmann distribution,
  with the  $\rho ({\bf x}) \propto \exp[-\beta \phi({\bf x})],$
  where $\phi$ is the gravitational potential related to the density $\rho$ by Poisson equation. This configuration, called isothermal sphere, has a  density profile $\rho \propto x^{-2}$ asymptotically.
Isothermal spheres with total energy $E$ and mass $M$, however,  cannot exist \cite{dlbtwo} if $(RE/GM^2) < -0.335$. Even when $(RE/GM^2)>-0.335$, the isothermal solution need not be stable. The stability of this solution can be investigated by studying the second variation of the entropy.
Such a detailed analysis shows that the following results are true:  (i) Systems with $(RE/GM^2)<-0.335$ cannot evolve into isothermal
spheres. Entropy has no extremum for such systems \cite{tppr}, \cite{dlbtwo}.
(ii) Systems with ($(RE/GM^2)>-0.335$) and ($\rho(0)> 709\,\rho(R)$) can
exist in a meta-stable (saddle point state) isothermal sphere configuration. Here $\rho(0)$ and $\rho(R)$ denote the densities at the center and edge respectively. The entropy extrema exist but they are not local maxima.
(iii) Systems with ($(RE/GM^2)> -0.335$) and ($\rho(0)<709\,\rho(R)$) can
form isothermal spheres, which are local maximum of entropy. These are striking peculiarities in the case of SMGS and seem to find application in the physics of globular clusters.

\section{Gravitational clustering in  an expanding background}  
 There is considerable amount of observational evidence to suggest that
  one of the dominant energy densities in the universe is contributed by self gravitating
  (nearly) point particles. The smooth average energy density of these particles drive
  the expansion of the universe while any small deviation from the homogeneous energy
  density will cluster gravitationally. It is often enough (and necessary) to use a statistical description and relate
  different statistical indicators (like the power spectra, $n$th order correlation functions, ....)
  of the resulting density distribution to the statistical parameters (usually the power spectrum) of the 
  initial distribution. 

The relevant scales at which gravitational clustering is nonlinear are less than
  about 10 Mpc,  while the expansion of the universe has a characteristic scale of about 4000
  Mpc \cite{cosmotext}. Hence, nonlinear gravitational clustering in an expanding universe can 
  be adequately described by Newtonian gravity  by introducing a {\it proper}
   coordinate for the $i-$th particle ${\bf r}_i$,
  related to the {\it comoving} coordinate  ${\bf x}_i$, by ${\bf r}_i = a(t) {\bf x}_i$ where
  $a(t)$  is the expansion factor. The Newtonian 
  dynamics works with the proper coordinates ${\bf r}_i$ which can be translated
  to the behaviour of the comoving coordinate ${\bf x}_i$ by this rescaling.

 If ${\bf x}(t,{\bf q} )$ is the position  of a  particle at time $t$ with  its initial position being ${\bf q}$, then equations for 
 gravitational clustering in an expanding universe, in the Newtonian
 limit, can be summarised as \cite{lssu}, \cite{tpiran}.

\begin{equation}
\ddot{\bf x} + { 2\dot a \over a} \dot{\bf x} = - {1 \over a^2} \nabla_{\bf x}
\phi;  \label{twnine}
\end{equation}
 
  \begin{eqnarray}
\ddot \phi_{\bf k} + 4 {\dot a \over a} \dot\phi_{\bf k}   &= & - {1 \over 2a^2} \int {d^3{\bf p} \over (2 \pi )^3} 
\phi_{ { {\bf k}\over2}+{\bf p}}  
\phi_{ { {\bf k }\over2}-{\bf p}}{\cal G}({\bf k},{\bf p}) \cr\nonumber \\
&+ &\lb{3H_0^2 \over 2}\rb  \int {d^3{\bf q} \over a} \lb{\bld k} . \dot {\bld x}\over k\rb ^2 e^{i{\bf k}.{\bf x}}, 
\label{powtransf} 
\end{eqnarray}
where $\bld x = \bld x(t, \bld q),
{\cal G}({\bf k},{\bf p})=(k/2)^2+p^2-2 ( \bld k . \bld p/ k)^2$
 and $\phi_{\bf k} (t)$ is the Fourier transform of the gravitational potential $\phi(t, {\bf x})$ due to perturbed density.

Equation (\ref{powtransf}) is exact but involves $\dot{\bf x}(t, \bld q)$  on the right hand side and hence, cannot be considered as closed.  Together, the two equations form a closed set but solving them exactly is an impossible task. It is, however, possible to
use this equation with several well motivated approximations \cite{tples} to obtain information about the system. I shall
briefly mention a few of them.

 Consider first the effect of a bunch of particles, in a virialized cluster, on the rest of the system. This
  is described, to the lowest order, by just the monopole moment of the cluster --
  which can be taken into account by replacing the cluster by a single particle at the centre of mass
  having appropriate mass. Such a replacement should not affect the evolution
at scales much bigger than the cluster size. At first sight, one may wonder how this feature (``renormalizability of gravity") is taken care
  of in equation (\ref{powtransf}). Inside a galaxy cluster, for example, the velocities
  ${\bf \dot x}$  can be quite high and one might wonder whether this could influence
  the evolution of $\phi_{\bf k}$ at all scales. This does not happen and, to the lowest order,
  the contribution from virialized bound clusters cancel \cite{tples}, \cite{lssu} in the two terms in the right hand side of (\ref{powtransf}).

In this limit, $\phi_{\bf k}$ is constant in time and the Poisson equation $-k^2\phi_{\bf k} = 4 \pi G \rho_ba^2\delta_{\bf k} \propto (\delta_{\bf k}/a)$ in the matter dominated universe with $a(t)\propto t^{2/3},\rho_b\propto a^{-3}$ implies that the density contrast  has the growing solution $\delta_{\bf k}(t) = [a(t)/a(t_i)] \delta_{\bf k}(t_i)$.  The power spectrum $P({\bf k},t) = \langle |\delta_{\bf k}(t)|^2\rangle$ and the correlation function
$\xi({\bf x},t)$ [which is the Fourier transform of the power spectrum] both grow as $a^2(t)$.
This  allows us to fix the evolution of clustering at sufficiently large scales
 uniquely. The clustering at these scales, 
which is well described by linear theory, grows as $a^2$.

 There is, however, an important caveat to this claim. While ignoring the  right hand side of (\ref{powtransf}) one 
   is comparing its contribution at any wave number ${\bf k}$ to the contribution in  linear theory. If at the relevant wavenumber, the contribution from linear evolution 
   is negligibly small, then
   the {\it only} contribution will come from the terms on the right hand side and, of course, we cannot
   ignore it in this case. This contribution will scale as 
   $k^2 R^2$, where $R$ is the typical scale of virialized systems and will lead to a development of $\delta_{\bf k} \propto k^2, P(k)\propto k^4$ at small $k$.  Thus, if  the large scales have too little
power intrinsically (i.e., if $n>4$ ), then
the long wavelength power will soon be dominated by the
$``k^4$ - tail'' of the short wavelength power arising from the
nonlinear clustering.  This is an interesting and curious result which is characteristic of gravitational
  clustering.
   
\section{Nonlinear scaling relations}

 As to be expected, cosmological expansion  completely changes the asymptotic nature of the problem. The problem has now become time dependent and it will be pointless to look for ``equilibrium solutions''. 
 
There are three key theoretical questions which are of considerable interest in this area which I will briefly summarise:  
  
\begin{itemize}
 \item
   If the initial power spectrum is sharply peaked in a narrow band of wavelengths, how does
  the evolution transfer the power to other scales? (This is, in some sense, analogous to determining
  the Green function for the gravitational clustering except that superposition will not work
  in the nonlinear context.)

\item
   Do the virialized structures formed in an expanding universe due to gravitational
  clustering have any invariant properties? Can their structure be understood from first principles?
\item
  Does the gravitational clustering at late stages wipe out the memory of initial conditions or does
  the late stage evolution depend on the initial power spectrum of fluctuations?
 
 \end{itemize}  
   
 To make any progress with these questions we need a  robust prescription which will  relate statistical indicators like the two-point correlation function in the nonlinear regime to the initial power spectrum. Fortunately, this problem
  has been solved \cite{hklm} to a large extent and hence, one can use this as a basis for attacking
 these questions.

The nonlinear mean
correlation function can be expressed in terms of the linear mean 
correlation function by the relation:
\begin{equation}
\bar \xi (a,x)=\cases{\bar \xi_L (a,l)&(for\ $  \bar
\xi<1$)\cr 
{\bar \xi_L(a,l)}^D &(for\ $ 1<\bar \xi<125$)\cr
11.7 {\bar \xi_L(a,l)}^{Dh/2} &(for\ $ 125<\bar
\xi$)\cr}\label{hamilton} 
\end{equation}
where $l = x [1 + \bar\xi(a,x)]^{1/D}$, $D=2,3$ is the dimension of space and $h$ is a constant. [The results of numerical simulation in 2D,  suggests that $h= 3/4$ asymptotically.  We will discuss the 3D results in
more detail below]. The numerical values are for $D=3$.
  
One could use this to examine whether the power spectrum (or the correlation function) 
   has a universal shape at late times, independent of initial power spectrum.
  This is indeed true \cite{jsbtp1} if the initial spectrum was sharply peaked. In this case,
   at length scales smaller than the initial scale at which the power is injected, the two point correlation 
   function has a universal asymptotic shape of $\bar \xi (a,x) \propto a^2x^{-1}(L+x)^{-1},$
   where $L$ is the length scale at which $\bar\xi \approx 200$. This can be understood as follows:

In the quasi-linear
phase, regions of high density contrast will undergo collapse and in the 
nonlinear phase
more and more virialized systems will get formed. 
We recall that, in the study of finite gravitating systems made of point particles and
interacting via Newtonian gravity, isothermal spheres play an important
role and are  the local maxima of entropy. Hence, dynamical
evolution drives the system towards an $(1/x^2)$ profile. Since, one expects
similar considerations to hold at small scales, during the late stages of evolution of the universe, we may hope that isothermal spheres with
$(1/x^2)$ profile may still play a role in the late stages of evolution of 
clustering in an expanding background. However, while converting the density profile to correlation function,
 we need to distinguish between two cases. 
In the quasi-linear regime, dominated by the collapse of high density peaks,
the density profile around any peak will scale as the correlation function and
we will have $\bar\xi\propto (1/x^2)$. On the other hand, in the nonlinear
end, we will be probing the structure inside a single halo and $\xi({\bf x}) $ 
will vary as $\langle \rho({\bf x + y}) \rho({\bf y}) \rangle$. If $\rho \propto |x|^{-\epsilon}$, then $\bar\xi \propto |x|^{-\gamma}$
with  $\gamma=2\epsilon -3$. This
gives $\bar\xi\propto (1/x)$ for $\epsilon=2$. Thus, if isothermal spheres
are the generic contributors, then we expect the correlation function to
vary as $(1/x)$ and nonlinear scales, steepening to $(1/x^2)$ at intermediate
scales. 
Further, since isothermal spheres are local maxima of entropy, a configuration like this should remain undistorted for a long duration. This
argument suggests that a $\bar\xi$ which goes as $(1/x)$ at small scales
and $(1/x^2)$ at intermediate scales is likely to  grow approximately as $a^2$ at all scales. 
At scales bigger than the scale at which power was originally injected,
   the spectrum develops a $k^4$ tail for reasons described before. This is confirmed by simulations for sharply peaked initial spectra. But if the initial spectrum is {\it not} sharply peaked, each band of power evolves by this
   rule and the final result is a lot messier.

  The second  question one could ask, concerns the density profiles of individual
  virialized halos. If the density field $\rho(a,{\bf x})$ at late stages  can 
be expressed as a superposition
of several halos, each with some density profile $f({\bf x})$ then 
the $i$-th halo centred at ${\bf x}_i$ will contribute a density
 $f({\bf x}-{\bf  x}_i,a)$  at the location ${\bf x}$. 
 The power spectrum for the 
density contrast, $\delta(a,{\bf x})=(\rho/\rho_b-1)$, will be $ P(k) = |f(k)|^2P_c(k)$,
  where $P_{\rm c}({\bf k},a)$
denotes the power spectrum of the distribution of centers of the halos. If the correlation function $\bar\xi \propto x^{-\gamma}$, the correlation function of the centres 
     $\bar\xi \propto x^{-\gamma_c}$ and the individual profiles are of the form 
     $f(x) \propto x^{-\epsilon}$, then this relation 
       translates to $  \epsilon = 3+(1/2)(\gamma - \gamma_c)
     $.  

At very nonlinear scales, the centres of the virialized clusters will coincide with 
     the deep minima of the gravitational potential. Hence, the power spectrum of the 
     centres will be proportional to the power spectrum of the gravitational potential 
     $P_\phi(k) \propto k^{n-4}$ if $P(k) \propto k^n$. Since the correlation functions
     vary as $x^{-(\alpha + 3)}$ when the power spectra vary as $k^\alpha$, it 
     follows that $\gamma = \gamma_c -4$. Substituting into the above relation,  we find
     that $\epsilon =1$ at the extreme nonlinear scales. On the other hand, in the 
     quasi-linear regime, reasonably large density regions will act as cluster
     centres and hence, one would expect $P_c(k)$ and $P(k)$ to scale 
     in a similar fashion. This leads to $\gamma \approx \gamma_c$, giving
     $\epsilon \approx 3$. So we would expect the halo profile to vary as $x^{-1}$ 
     at small scales steepening to $x^{-3}$ at large scales. A simple interpolation
     for such a density profile will  be 
     \begin{equation}
     f(x) \propto {1\over x(x+l)^2}.
\label{pronwf}
     \end{equation}
     Such a profile, usually called NFW profile \cite{NFW}, is often used in cosmology.   
The argument given above, however, is very tentative and it is difficult to obtain (\ref{pronwf})
from a more rigorous theoretical analysis.

In fact, it is possible to reach different conclusions regarding the asymptotic
evolution of the system from different physical assumptions \cite{tpsunu}.
The NSR in (\ref{hamilton}) for 3-D with constant $h$ leads to the asymptotic correlation function
\begin{equation}
\bar\xi(a,x)\propto a^{\frac{2 \gamma}{n+3}} x^{-\gamma};\qquad
\gamma=\frac{3 h (n+3)}{2+h(n+3)}
\label{genres}
\end{equation}
 for an initial
spectrum which is scale-free power law with index $n$.
If we assume that the
evolution gets frozen in proper coordinates at highly nonlinear scales then it is easy to show that $h=1$.
{\it If} this assumption (called stable clustering) {\it is} valid, then the late time  behaviour of $\xb$ is
strongly dependent on the  initial conditions and (\ref{genres}) shows
 that $\xb$ at nonlinear  scales will be as,
\begin{equation}
\bar\xi(a,x) \propto a^{\frac{6}{n+5}} x^{-\frac{3(n+3)}{n+5}};\qquad (\bar\xi 
\gg 200).
\end{equation}
 In other words the two (apparently reasonable) requirements:
(i) validity of stable clustering at highly nonlinear scales and
(ii) the independence of late time behaviour from initial conditions, 
{\it are mutually
exclusive}.  [At present, there exists some evidence 
from simulations \cite{tpjpo} that 
this process, called stable clustering, does {\it not} occur in 
the $a\propto t^{2/3}$ cosmological model; but this result is not definitive].  

In the very nonlinear limit, the correlation function probes the interiors of individual
halos and we have $ \epsilon=(1/2)(3+\gamma)$. [This corresponds to $P_c=$ constant,
$\gamma_C=3$
in the previous discussion.] If $\gamma$ depends on $n$ so will $\epsilon$ and the individual
halos will remember the initial power spectrum.

We  can obtain
a $\gamma$  which is independent of initial power law index provided
$h$ satisfies the condition $h(n+3)=c$, a constant.  In this case, the halo profile will be given by
$
\epsilon=3 (c+1)/(c+2). 
$
Note that we are now demanding the asymptotic value of $h$ to {\it explicitly 
depend} on the initial conditions though the {\it spatial} dependence of $\xb$ 
does not.
As an example of the power of such a --- seemingly simple --- analysis, note the 
following: Since $c \geq 0 $, it follows that $\epsilon > (3/2)$; invariant 
profiles
with shallower indices (for e.g with $\epsilon=1$) discussed above are not consistent 
with the evolution described above. One requires very high resolution simulations to 
verify the condition $h(n+3)=c$ and the current results are inconclusive.

     \end{document}